\documentclass[doublecol]{epl2}
\usepackage{amssymb,amsmath,graphicx,setspace,color,rotating,subfigure,url}
\usepackage{extarrows}
\title{Network analysis of the worldwide footballer transfer market}

\author{Ming-Xia Li\inst{1,2} \and Wei-Xing
  Zhou\inst{2,3,4}\footnote{e-mail: wxzhou@ecust.edu.cn} \and H. Eugene
  Stanley\inst{5}}
\shortauthor{M.-X. Li \etal}

\institute{
  \inst{1} Research Institute of Sports Economics, East China University of Science and Technology, Shanghai 200237, China\\
  \inst{2} Research Center for Econophysics, East China University of
  Science and Technology, Shanghai 200237, China\\
  \inst{3} Department of Finance, East China University of Science and
Technology, Shanghai 200237, China\\
  \inst{4} Department of Mathematics, East China University of Science
  and Technology, Shanghai 200237, China\\
  \inst{5} Center for Polymer Studies and Department of Physics, Boston
  University, Boston, Massachusetts 02215, USA\\
}

 \pacs{89.20.-a}{Interdisciplinary applications of physics}
 \pacs{01.80.+b}{Physics of games and sports}
 \pacs{89.75.Da}{Systems obeying scaling laws}

\abstract{The transfer of football players is an important part in football games. Most studies on the transfer of football players focus on the transfer system and transfer fees but not on the transfer behavior itself. Based on the 470,792 transfer records from 1990 to 2016 among 23,605 football clubs in 206 countries and regions, we construct a directed footballer transfer network (FTN), where the nodes are the football clubs and the links correspond to the footballer transfers. A systemic analysis is conduced on the topological properties of the FTN. We find that the in-degrees, out-degrees, in-strengths and out-strengths of nodes follow bimodal distributions (a power law with exponential decay), while the distribution of link weights has a power-law tail. We further figure out the correlations between node degrees, node strengths and link weights. We also investigate the general characteristics of different measures of network centrality. Our network analysis of the global footballer transfer market sheds new lights into the investigation of the characteristics of transfer activities.}

\begin{document}
\maketitle

\section{Introduction}
\label{S1:Introduction}

Football matches are widely regarded as the most influential sport in the world. According to a survey of FIFA in 2006, 4\% of the world's population are actively involved in football. The transfer of football players became part of the football game after 1893. Up till now, the transfer records were updated again and again, and the news of footballer transfers was listed on the headlines of sports news numerous times. The active football market needs to be constantly stirred, and the continually sensational transfer is one of the driving forces.

From the birth of the world's first million transfer to the free transfer of European football, the beautiful sport has become a booming industry. However, research on the transfer of football players focuses mostly on the transfer system and transfer fees and there is less research on the transfer behavior itself. It is found that the transfer fee is different in various segments in English professional football sports \cite{Dobson-Gerrard-1999-JSM}, and the determination of transfer fee is similar in both professional and nonleague football sports \cite{Dobson-Gerrard-Howe-2000-AE}. Usually football players benefit from the transfer system \cite{Dietl-Franck-Lang-2008-EJLE}; However the transfer system might also obstruct the free movement of football players between Member States in Europe and may restrict the ability of most clubs to compete for elite players \cite{Pearson-2015-ELJ}. It is also found that in the English Football League there is no racial difference in footballers' transfer prices \cite{Medcalfe-2008-AEL}.

In the era of big data, social science researchers consider social network analysis as an important tool to understand and excavate empirical laws of social behavior \cite{Barabasi-2005-Nature,Jiang-Xie-Li-Podobnik-Zhou-Stanley-2013-PNAS,Gonzalez-Hidalgo-Barabasi-2008-Nature}. With the rapid development of information transmission and storage technology, lots of human daily activities have been recorded. In the face of massive log data of human behavior, computational social science has attracted wide attention of researchers in recent years \cite{Wasserman-Faust-1994,Squartini-Lelyveld-Garlaschelli-2013-SR,Bargigli-Iasio-Infante-Lillo-Pierobon-2014-QF,Li-Palchykov-Jiang-Kaski-Kertesz-Micciche-Tumminello-Zhou-Mantegna-2014-NJP}. Barab\'{a}si points out that the ideas and methods of complex networks in the process of understanding complex social phenomena will be indispensable tools. The network links between the two bodies follow different dynamic laws during the construction process, such as assortative, correlation, and proximity \cite{Wasserman-Faust-1994,Jiang-Zhou-2010-PA,Li-Palchykov-Jiang-Kaski-Kertesz-Micciche-Tumminello-Zhou-Mantegna-2014-NJP}.

In recent years, network analysis has been applied to study different networks constructed from variant attributes of the football market, such as the bipartite network of players and clubs \cite{Onody-deCastro-2004-PRE}, football passing networks among players \cite{Yamamoto-Yokoyama-2011-PLoS1,Clemente-Martins-Kalamaras-Wong-Mendes-2015-IJPAS,Clemente-Martins-Wong-Kalamaras-Mendes-2015-IJPAS,Mclean-Salmon-Gorman-Stevens-Solomon-2018-HMS,Mendes-Clemente-Mauricio-2018-JHK}, zone-specified passing networks \cite{Cotta-Mora-Merelo-MereloMolina-2013-JSSC,Narizuka-Yamamoto-Yamazaki-2014-PA,Clemente-Martins-Mendes-2016-Kinesiology,Clemente-Jose-Oliveira-Martins-Mendes-Figueiredo-Wong-Kalamaras-2016-JHSE}, directed footballer transfer networks (FTNs) \cite{Liu-Liu-Lu-Wang-Wang-2016-PLoS1}, and mutual footballer transfer networks (MFTNs)  \cite{Li-Xiao-Wang-Zhou-2018-IJMPB}. In this Letter, we construct a direct transfer network to investigate the features of the transfer events of football players between different football club. Other than the study of Liu et al. \cite{Liu-Liu-Lu-Wang-Wang-2016-PLoS1}, we mainly focus on the topological properties of the FTN.

\section{The footballer transfer network}
\label{S1:FTN}

The football player transfer records from 1990 to 2016 were retrieved from http://www.transfermarkt.com. There are 470,792 transfers among 23,765 worldwide football clubs.

We construct a footballer transfer network (FTN), in which the nodes refer to the clubs and a directed link $i\to{j}$ forms when a player is transferred from club $i$ toclub $j$. The FTN is composed of 23,765 nodes and 243,770 directed links. It is possible that there are multiple transfers from one club to another club. In this case, only one directed link is drawn and its weight is the number of transfers. In contrast, the FTN analyzed by Liu et al. contains 410 nodes and 6316 directed links and the time period is from 2011 to 2015 \cite{Liu-Liu-Lu-Wang-Wang-2016-PLoS1}.

\begin{figure}[tb]
\centering
  \includegraphics[width=0.9\linewidth]{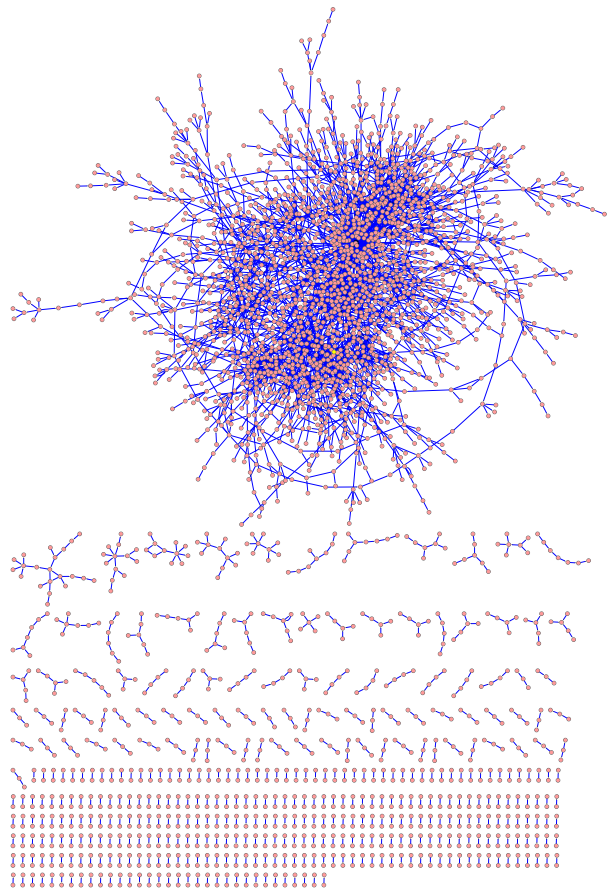}
  \caption{\label{fig:FTN:1day} A sample of the daily FTN construct from about 4000 transfer records on 1 January 2007. Each node refer to a football club in the world, and a direct link corresponds to the transfer events occurred from the original club to the target club on 1 January 2007.}
\end{figure}

There are 39 connected components or sub-networks in the FTN. The largest component contains 23,669 nodes (99.6\% of all nodes). Each smaller component has no more than 11 nodes. The components are dominated by trees and the largest network is very sparse. Hence the density of the whole network is very small (close to 0.0004) and the average clustering coefficient is not large (about 0.20). A sample FTN constructed from about 4000 transfer records on 1 January 2007 is illustrated in fig.~\ref{fig:FTN:1day}.



\begin{figure*}[htb]
  \centering
  \includegraphics[width=0.31\linewidth]{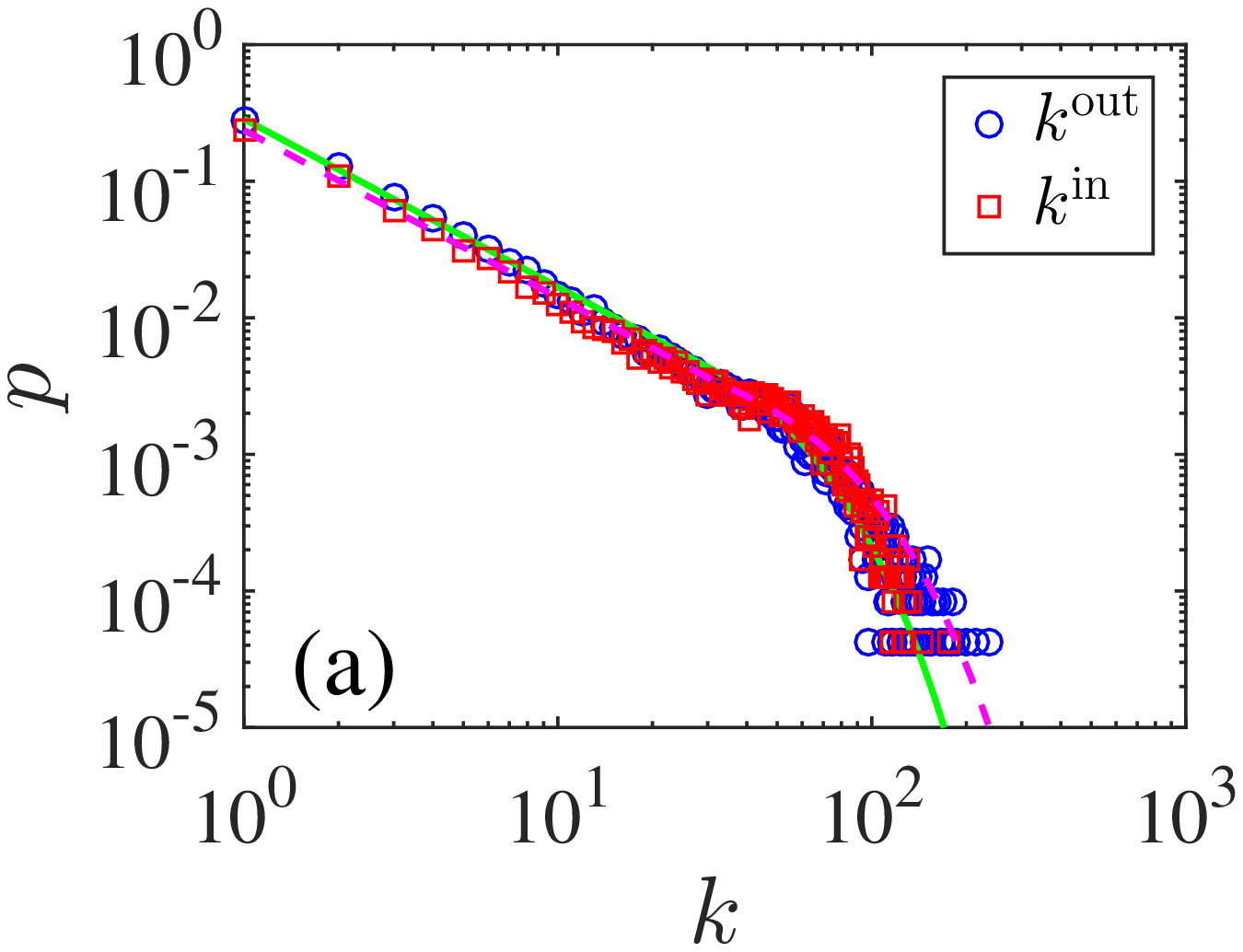}
  \includegraphics[width=0.31\linewidth]{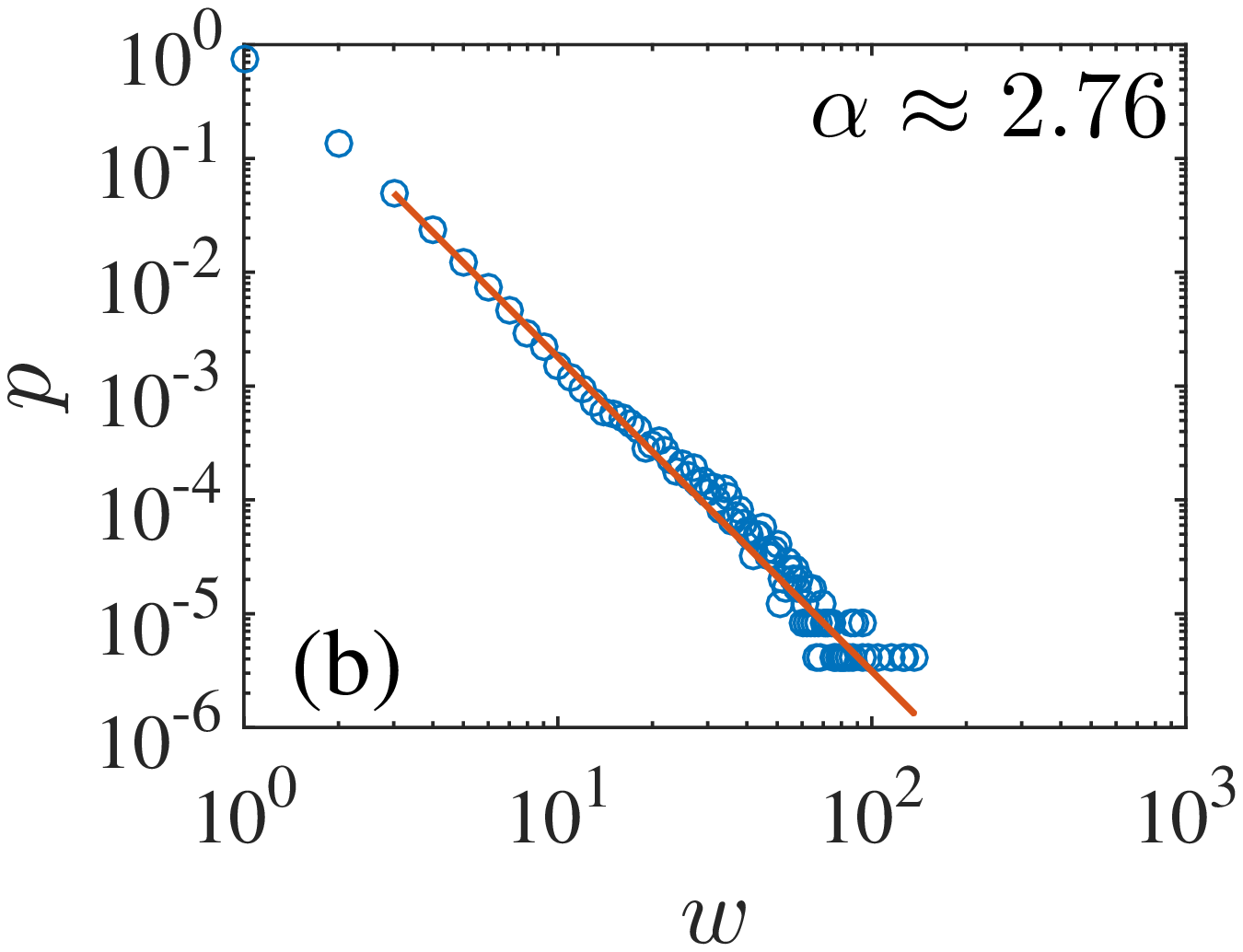}
  \includegraphics[width=0.31\linewidth]{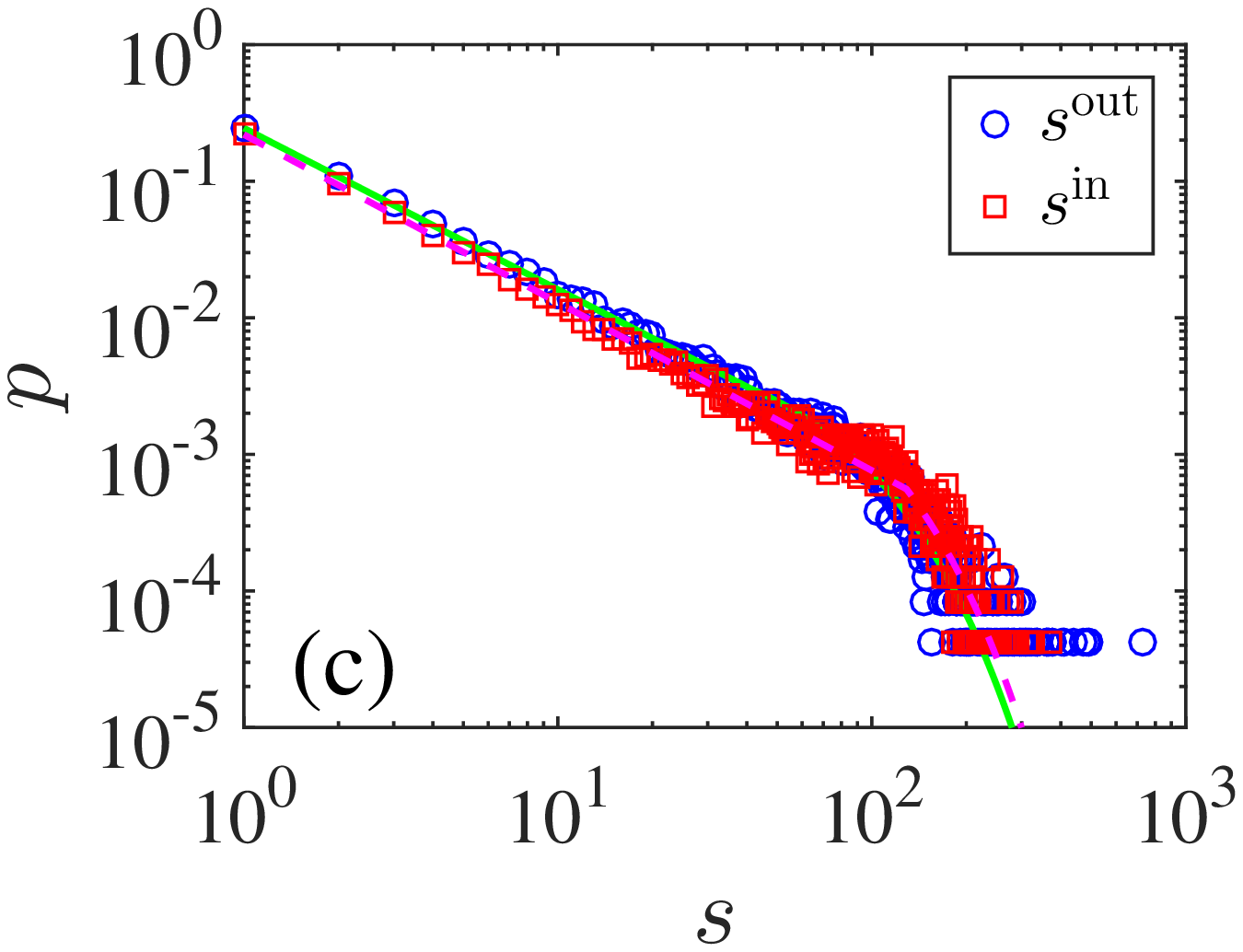}
  \caption{\label{fig:pdf} Probability density distribution. (a) Distributions of the out-degrees $k^\text{out}$ and the in-degrees $k^\text{in}$. The green solid line and the magenta dashed line are the fits to the bimodal distributions. (b) Distribution of link weights $w$. The red dashed line is the fitted power law. (c) Distributions of the out-strengths $s^\text{out}$ and the in-strengths $s^\text{in}$. The green solid line and the magenta dash line are the corresponding fits to the bimodal distributions.}
\end{figure*}

\section{Results}

\subsection{Distributions}


In the FTN, a club's out-degree $k^\text{out}_i$ is the number of clubs that received its football players, and a club's in-degree $k^\text{in}$ is the number of clubs which transferred football players to it.
We find that the average node degrees, $\langle k^\text{out} \rangle$ and $\langle k^\text{in} \rangle$, are close to 10.26.
The maximum out-degree and in-degree are respectively $k^\text{out}_{\max} = 235$ and $k^\text{in}_{\max} = 177$, corresponding to the same football club, ``Parma (Italy)'', which currently competes in Series A.

We find that there are no nodes with $k^\text{in} = 0$ and $k^\text{out} = 0$, meaning that no club is isolated from other clubs in the transfer market.
The number of nodes with $(k^\text{in} = 0$ and $k^\text{out} > 0)$ is 4746. These nodes usually correspond to football training clubs. For example, club ``Yonsei Univ (Korea, South)'', a squad of Yonsei University, has and in-degree of 0 and an out-degree of 24. A rookie football player goes to these clubs for training because professional clubs do not welcome inexperienced players. They need to improve their skills and show their abilities on football pitches.
The number of nodes with $k^\text{in} > 0$ and $k^\text{out} = 0$ is 1779. These nodes most probably correspond to professional football clubs. For example, club ``Reno FC (United States)'',  a second tier in United Soccer League, has an in-degree of 18 and an out-degree of 0. These clubs may have good rankings and provide good salaries. Most football players want to enter those clubs and will scarcely transfer out.


In the FTN, the weight $w_{ij}$ of a directed link $i{\to}j$ represents the number of football players transfered from club $i$ to club $j$. The average weight of all the links in the FTN is $\langle w \rangle = 1.67$, indicating the average number of football players transferred from one club to another is less than 2. It implies that most football clubs have transferred only one football player. The link with the maximum weight connects two football clubs, `` Akademia FCSM (Russia)'' and ``Spartak Moskow II (Russia)''. The reason is that the former club is the academy of the next one.

The in-strength and out-strength of a node $i$ can be calculated as follows:
\begin{equation}
s^\text{out}_i= \sum_{j\in\mathbb{N}} w_{ij}~~{\textrm{and}}~~ s^\text{in}_i= \sum_{j\in\mathbb{N}} w_{ji}.
\end{equation}
We find the average node strengths are $\langle s^\text{out} \rangle = \langle s^\text{in} \rangle = 17.17$. The maximum node strengths are respectively $s^\text{out}_{\max} = 724$ and $s^\text{in}_{\max} = 371$, which correspond again to club ``Parma (Italy)''.

Fig.~\ref{fig:pdf} shows the distributions of node degrees $k$, link weights $w$ and node strengths $s$. One can observe the similar distribution shapes for node degrees and node strengths in fig.~\ref{fig:pdf}(a) and in fig.~\ref{fig:pdf}(c), which can be fitted well by a bimodal distribution \cite{Wu-Zhou-Xiao-Kurths-Schellnhuber-2010-PNAS},
\begin{equation}
	p(x)=\left\{
	\begin{array}{ll}
		x^{-\gamma{\color{red}{-1}}}, 	& x<x_0\\
		e^{-\beta x},	& x>x_0
	\end{array}
	\right..
\end{equation}
where $x_0$ is the separate point of bimodal distribution. The fitted parameters are listed in Table \ref{TB:PDF:k:s}, in which we also present the results for the MFTN \cite{Li-Xiao-Wang-Zhou-2018-IJMPB}. It shows that although the FTN and the MFTN have the same degree and strength distributions qualitatively, they have quantitative differences. For the degree distribution, the exponential tail for the FTN decays faster than the MFTN. For the strength distribution, the power-law part for the FTN decays faster than the MFTN. Low-degree clubs are inclined to transfer with high-degree clubs in order to gain more income and this preferential attachment mechanism leads to the power law part of the degree distribution \cite{Barabasi-Albert-1999-Science}. In contrast, high-degree clubs usually do not has such a significant tendency and such a somewhat random transfer mechanism results in the exponential part of the degree distribution.

\begin{table}[htb]
  \centering
  \caption{\label{TB:PDF:k:s} Estimated parameters of the bimodal distributions of the directed footballer transfer network (first two rows) with a comparison to the mutual transfer network (third row) \cite{Li-Xiao-Wang-Zhou-2018-IJMPB}.}
  \smallskip
  \begin{tabular}{llllllllllllllll}
    \hline\hline
         & \multicolumn{1}{c}{$\gamma$} & \multicolumn{1}{c}{$\beta$} & & & \multicolumn{1}{c}{$\gamma$} & \multicolumn{1}{c}{$\beta$} \\
    \hline
    $k^\text{out}$  & 0.23 & 0.04  &&  $s^\text{out}$ & 0.18 & 0.02 \\
    $k^\text{in}$   & 0.23 & 0.03  &&  $s^\text{in}$  & 0.23 & 0.02 \\
    $k$             & 0.24 & 0.07  &&  $s$            & 0.04 & 0.02 \\
    \hline\hline
  \end{tabular}
\end{table}

The distribution of link weights in fig.~\ref{fig:pdf}(b) can be approximated by a power-law distribution:
\begin{equation}
	\begin{array}{ll}
		p(w) = w^{-\alpha} & {\mathrm{for}}~ w>w_{\min}.
	\end{array}		
\end{equation}
Using the method of Clauset et al. \cite{Clauset-Shalizi-Newman-2009-SIAMR}, we obtain that $\alpha = 2.76$ and $w_{\min} = 2$. The link weight distribution for the MFTN also follows a power-law tailed distribution, in which $\alpha = 2.40$ and $w_{\min} = 8$ \cite{Li-Xiao-Wang-Zhou-2018-IJMPB}. Again, the distribution for the FTN decays faster than the MFTN.

\begin{figure*}[htb]
\centering
  \includegraphics[width=0.32\linewidth]{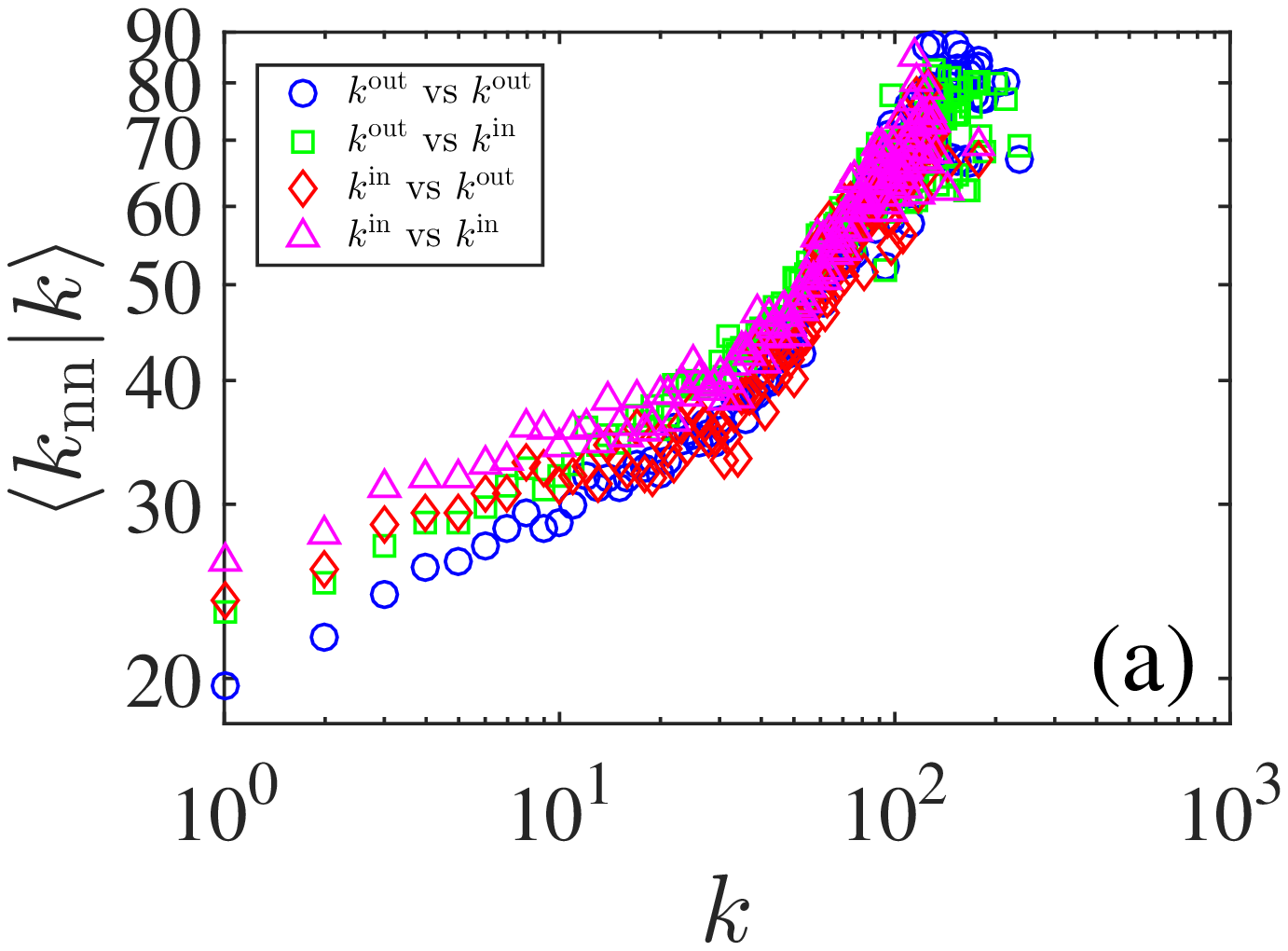}
  \includegraphics[width=0.32\linewidth]{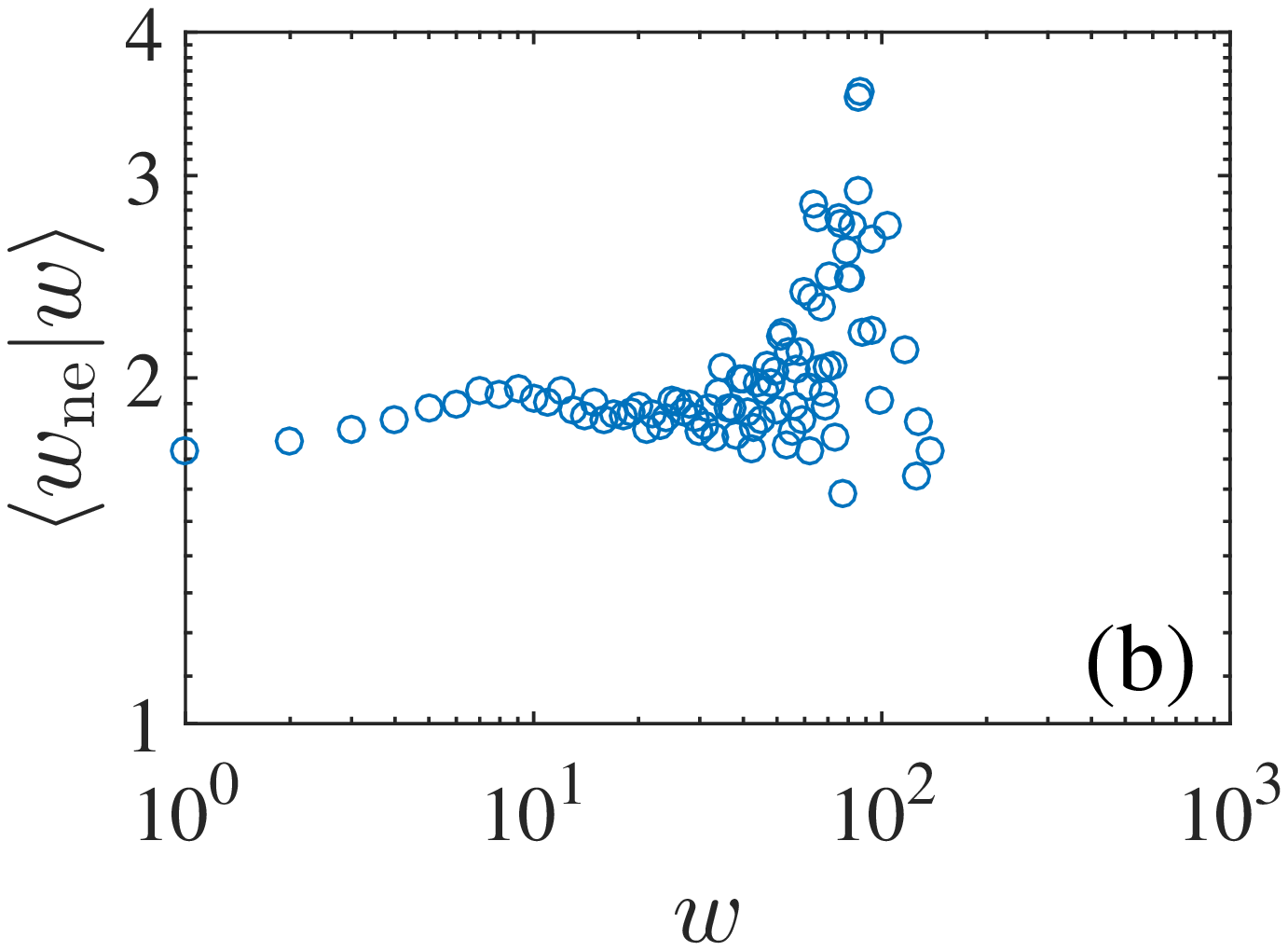}
  \includegraphics[width=0.32\linewidth]{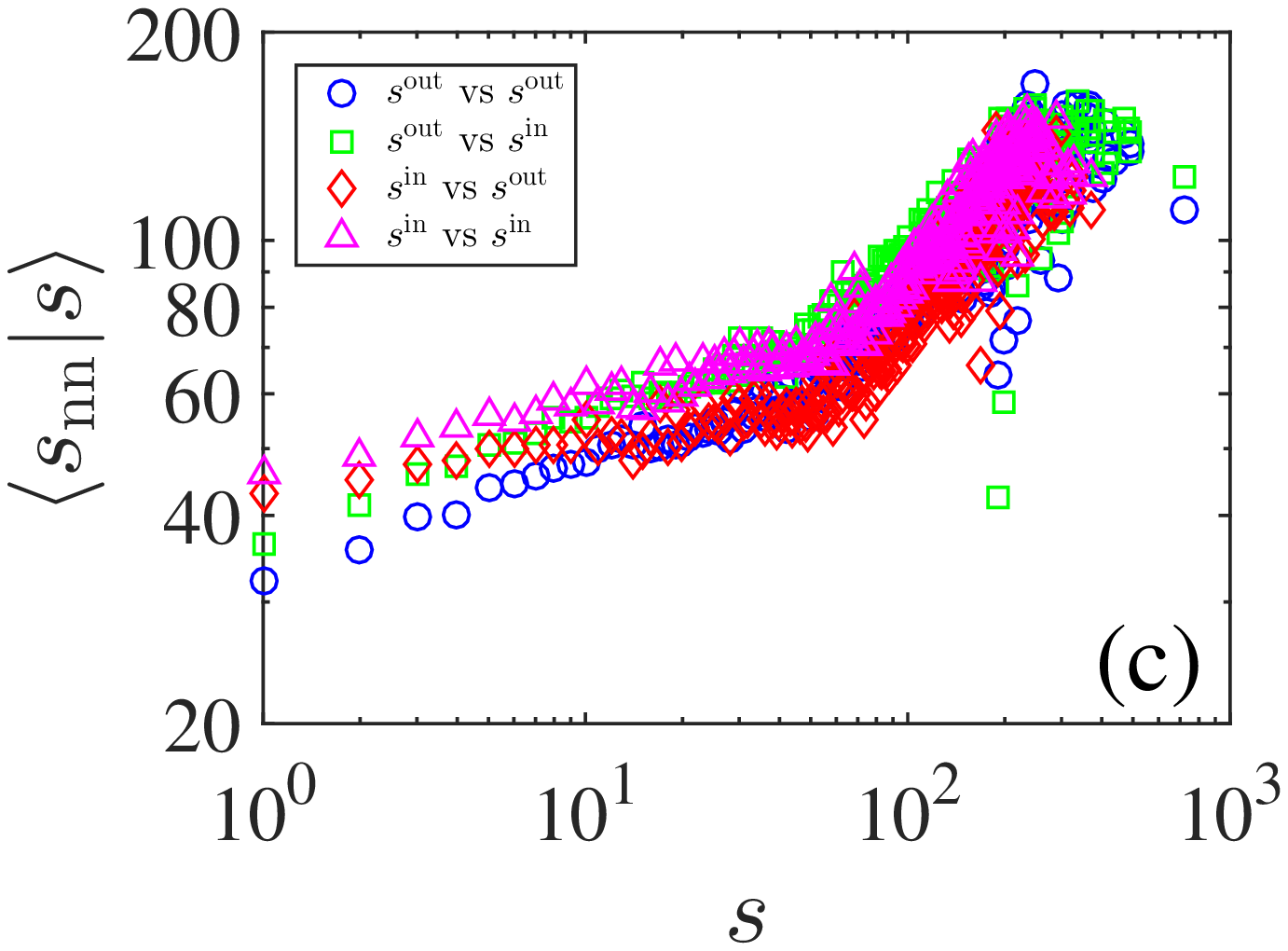}
  \caption{\label{fig:corr:1} Assortative mixing. (a) Average neighbor node degree $\langle k_\text{nn} | k \rangle$ as a function of degree $k$. Four curves correspond to four correlations between node out-degree $k^\text{out}$ and node in-degree $k^\text{in}$. For example, $k^\text{out}$ vs $k^\text{in}$ is the average neighbor node in-degree $\langle k^\text{in}_\text{nn} | k^\text{out} \rangle$ as a function of out-degree $k^\text{out}$. (b) Average neighbor link weight $\langle w_\text{ne} | w \rangle$ as a function of weight $w$. (c) Average neighbor node strength $\langle s_\text{nn} | s \rangle$ as a function of strength $s$.}
\end{figure*}

\subsection{Correlations}

In social networks, an interesting result is that nodes with similar characteristics will prefer to connect with each other, called assortative mixing \cite{Newman-2002-PRL}. In some case of social networks, it is also known as the homophily \cite{McPherson-SmithLovin-Cook-2001-ARS,Currarini-Jackson-Pin-2009-Em,Kovanen-Kaski-Kertesz-Saramaki-2013-PNAS,Kossinets-Watts-2009-AJS}. One way to detect such feature in FTN is to investigate the average neighbor node degree:
\begin{equation}
k_{{\rm{nn}},i} = \frac{1}{k_i} \sum_{j\in \mathcal{N}_i} k_j,
\label{eq:knn}
\end{equation}
where $\mathcal{N}_i$ is the set of nearest neighbor nodes of $i$ which are directly connected to node $i$. By averaging $k_{\text{nn}}$ over all the nodes with degree $k$, we can observe the correlation between $k_\text{nn}$ and $k$, as shown in fig.~\ref{fig:corr:1}(a). Since a node has an out-degree and an in-degree, we obtain four correlations between node degree $k$ and average neighbor node degree $k_\text{nn}$. All the four lines have an increasing trend with the increase of node degree. It indicates that the FTN exhibits an assortative mixing pattern, just like many other social networks \cite{Newman-2002-PRL,Catanzaro-Caldarelli-Pietronero-2004-PRE,Rivera-Soderstrom-Uzzi-2010-ARS,Li-Jiang-Xie-Micciche-Tumminello-Zhou-Mantegna-2014-SR}. The direction of links in the FTN do not have great impact on this feature \cite{Li-Xiao-Wang-Zhou-2018-IJMPB}. Besides that, each curve can be approximated by two power laws.

The average neighbor edge weight $w_{{\rm{ne}},ij}$ of link $i{\to}j$ for the MFTN has been studied \cite{Li-Xiao-Wang-Zhou-2018-IJMPB}. We can calculate $w_{{\rm{ne}},ij}$ for the FTN similarly. We create a new undirected subordinate network $\text{FTN}^*$, where the nodes are mapped from the links in the FTN and a link is created if two links $i{\to}j$ (or $j{\to}i$) and $i{\to}k$ in the FTN have a common node $i$. For simplicity, we do not consider the bi-directed links as a pair of neighbor links, that is, we require that $j\neq k$. Thus, the neighbors of a link in the FTN is converted to the neighbors of a node in the $\text{FTN}^*$. The average neighbor link weight $w_{{\rm{ne}},ij}$ of a link is calculated as follows:
\begin{equation}
\begin{aligned}
w_{{\rm{ne}},ij} &= \frac{\sum\limits_{a\neq i,b\neq j}(w_{aj}+w_{ib}+w_{ja}+w_{bi})}{k^\text{out}_i+k^\text{in}_i+k^\text{out}_j+k^\text{in}_j-4}\\
& = \frac{s^\text{out}_i+s^\text{in}_i+s^\text{out}_j+s^\text{in}_j-2w_{ij}-2w_{ji}}{k^\text{out}_i+k^\text{in}_i+k^\text{out}_j+k^\text{in}_j-4}.
\end{aligned}
\end{equation}
The average neighbor link weight $\langle w_{{\rm{ne}}} | w \rangle$ for the edges of weight $w$ is calculated by averaging $w_{{\rm{ne}},ij}$ over those edges of weight $w$.
Fig.~\ref{fig:corr:1}(b) shows the average neighbor link weight $\langle w_{\rm{{\rm{ne}}}} | w\rangle$ as a function of weight $w$. $\langle w_{\rm{{\rm{ne}}}} | w \rangle$ moves within $[1.7, 2]$ when $w < 30$. When $w > 30$, the fluctuation of $\langle w_{\rm{{\rm{ne}}}} | w \rangle$ become drastic. It indicates that the number $w$ of transfers between two clubs in the FTN distributes more randomly than the number of clubs that a club transferred with $k$.

The average nearest neighbor strength $s_\text{nn}$ of node $i$ is the average strength of $i$'s nearest neighbors:
\begin{equation}
s_{{\rm{nn}},i} = \frac{1}{k_i} \sum_{j\in \mathcal{N}_i} s_j.
\label{eq:snn}
\end{equation}
Like the similarity between the distributions of node degrees and node strengths (see fig.~\ref{fig:pdf}), the correlation between $s_\text{nn}$ and $s$ is quite similar to the correlation between $k_\text{nn}$ and $k$.

Since the distributions of node degrees and node strengths are similar to each other, one may expect that the average node strength has a linear correlation with the node degree \cite{Barrat-Barthelemy-PastorsSatorras-Vespignani-2004-PNAS,Onnela-Saramaki-Hyvonen-Szabo-Menezes-Kaski-Barabasi-Kertesz-2007-NJP},
\begin{equation}
\langle s | k \rangle \sim k^{\alpha_{ks1}}
\end{equation}
with $\alpha_{ks1} = 1$, if there is no correlation between the node degree and the weights of the edges linked to the node. We also check the correlation between the product of node strength and node degree,
\begin{equation}
\langle s_is_j | k_ik_j \rangle \sim (k_ik_j)^{\alpha_{ks2}}
\end{equation}
where nodes $i$ and $j$ belong to link $i{\to}j$.

\begin{figure}[htb]
\centering
  \includegraphics[width=0.9\linewidth]{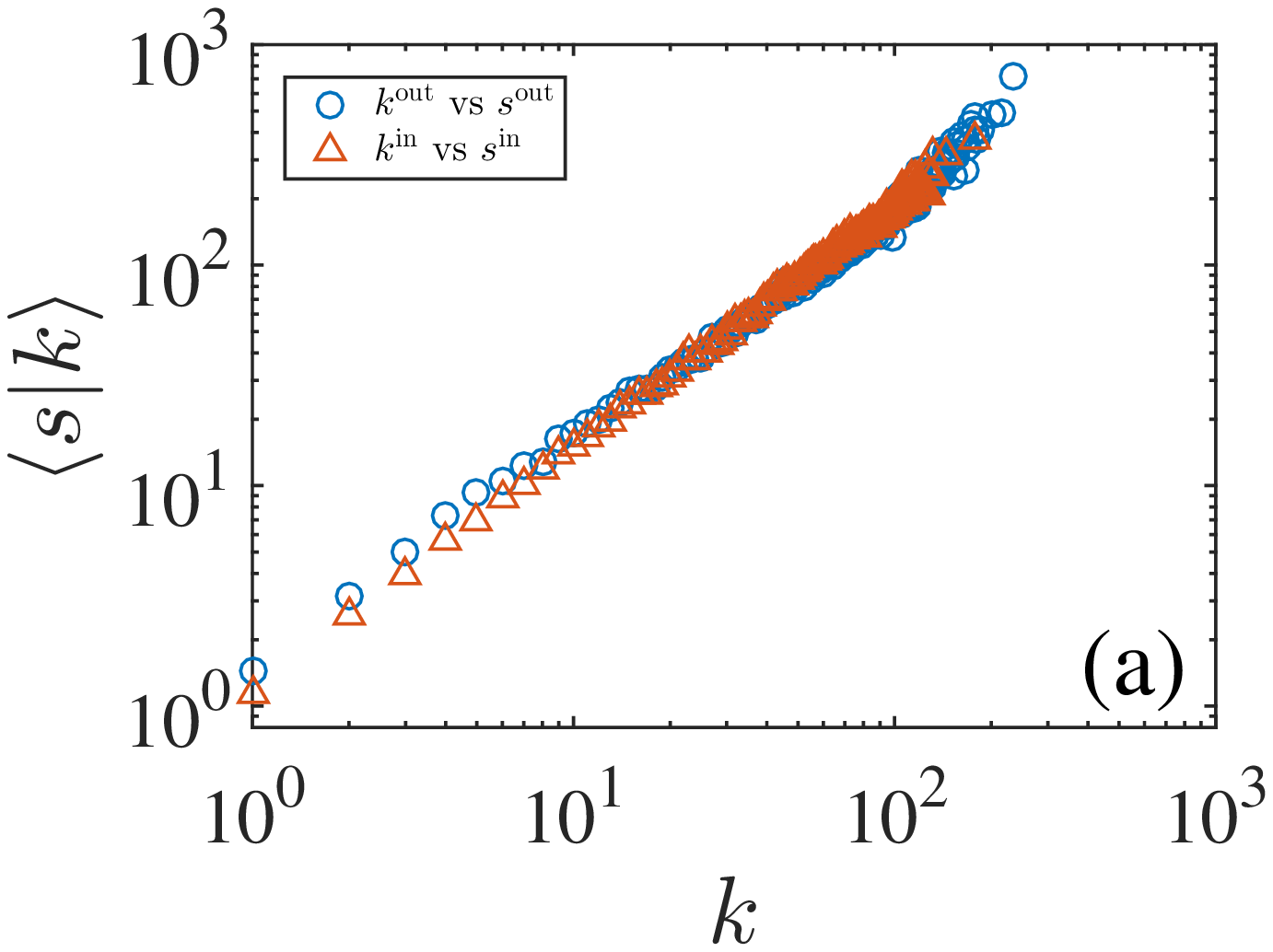}
  \includegraphics[width=0.9\linewidth]{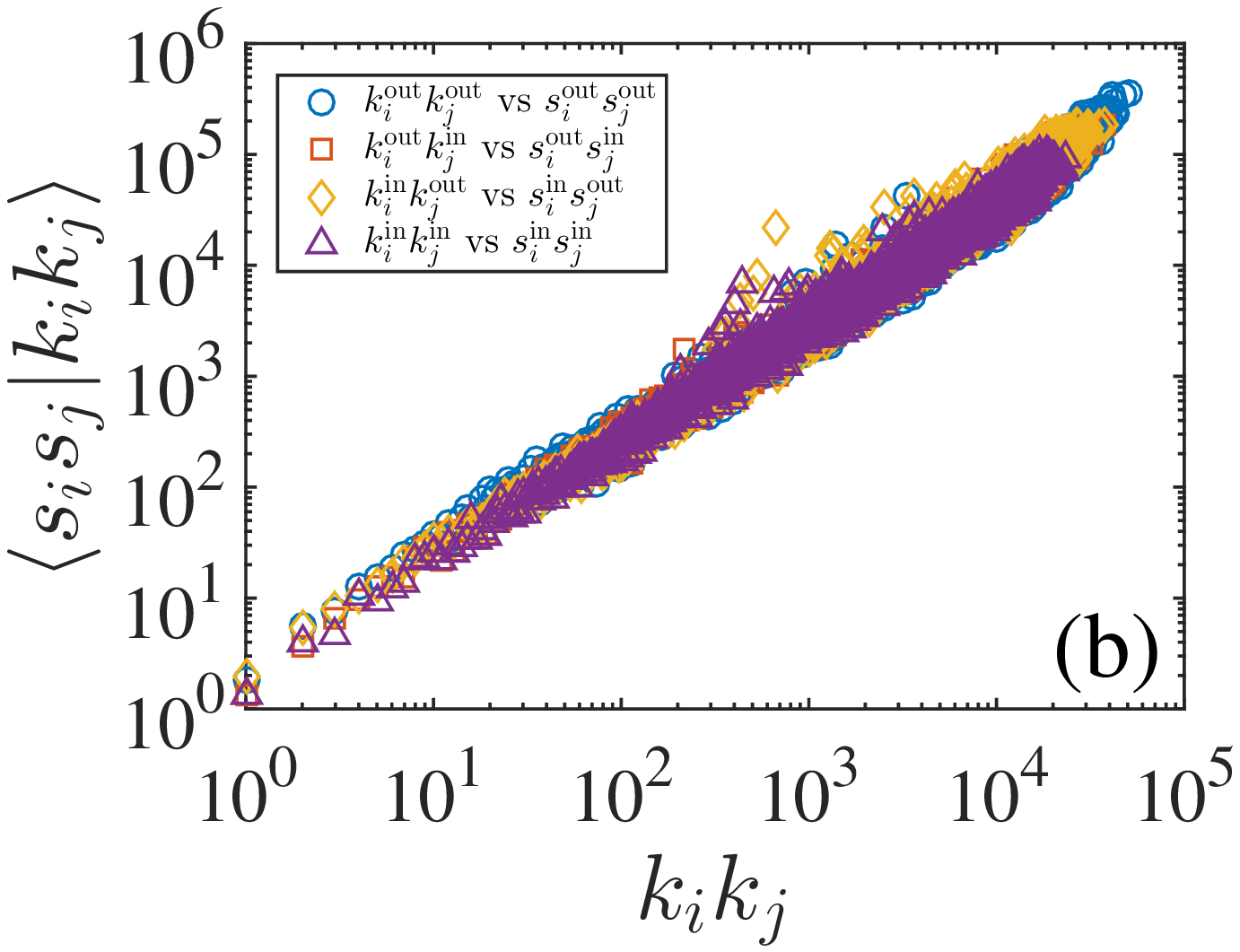}
  \caption{\label{fig:corr:2} Correlations between node degrees and node strengths. (a) The average strength $\langle s_i \rangle$ of the nodes with degree $k$. (b) The average strength product $\langle s_is_j \rangle$ of the directed links from node $i$ to node $j$ whose degrees are $k_i$ and $k_j$.}
\end{figure}

Fig.~\ref{fig:corr:2} shows the correlations between node degrees and node strengths. Very nice power-law dependence is observed. The power-law exponents are estimated as follows: $\alpha_{ks1}(\text{out}) = 1.06$,  $\alpha_{ks1}(\text{in}) = 1.10$ in fig.~\ref{fig:corr:2}(a) and $\alpha_{ks1}(\text{out,out}) = 1.08$,  $\alpha_{ks1}(\text{out,in}) = 1.09$, $\alpha_{ks1}(\text{in,out}) = 1.10$,  $\alpha_{ks1}(\text{in,in}) = 1.07$ in fig.~\ref{fig:corr:2}(b).
All the values of $\alpha_{ks1}$ and $\alpha_{ks2}$ are greater than 1, which are close to the case of MFTN \cite{Li-Xiao-Wang-Zhou-2018-IJMPB}.

\begin{figure}[htb]
\centering
  \includegraphics[width=0.9\linewidth]{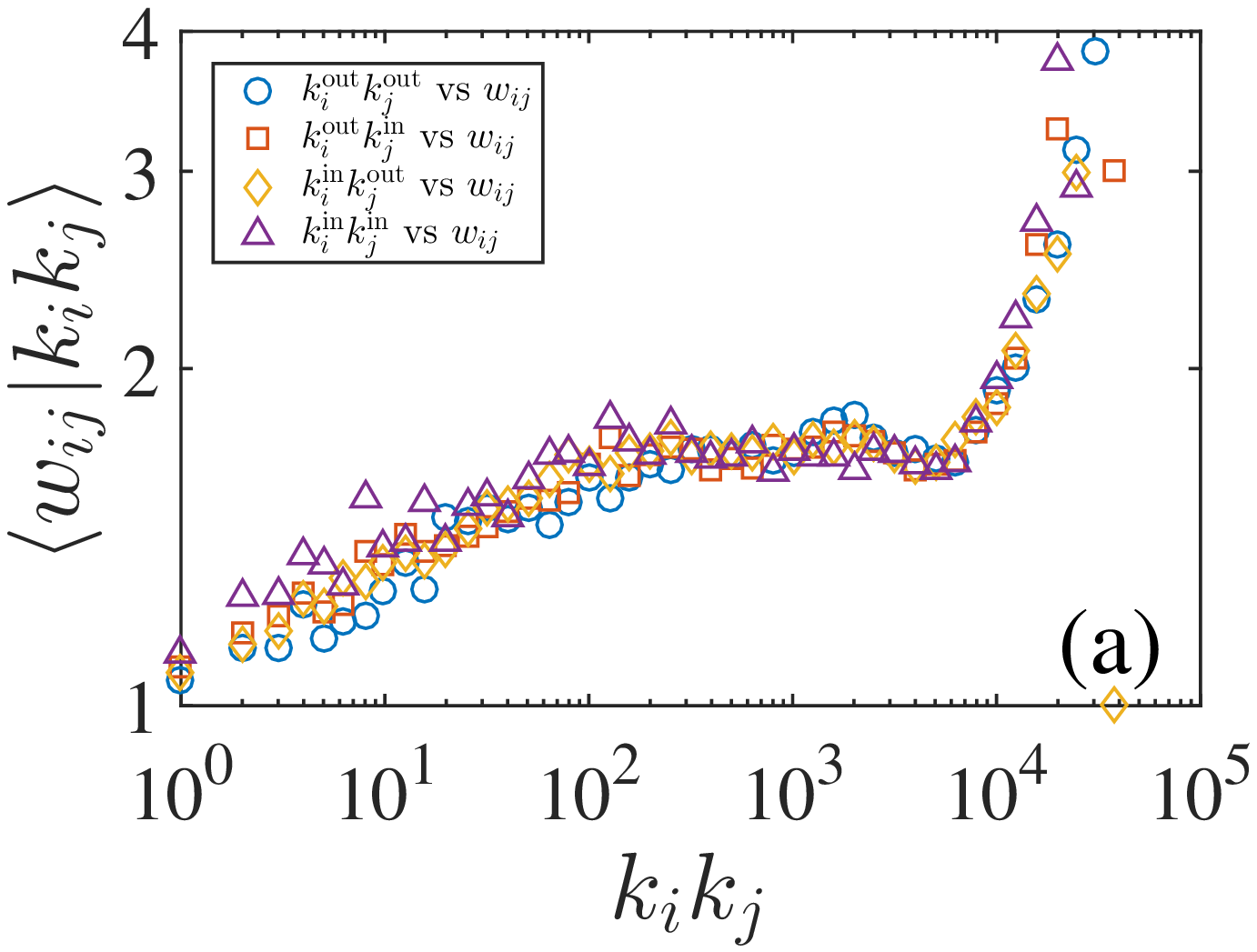}
  \includegraphics[width=0.9\linewidth]{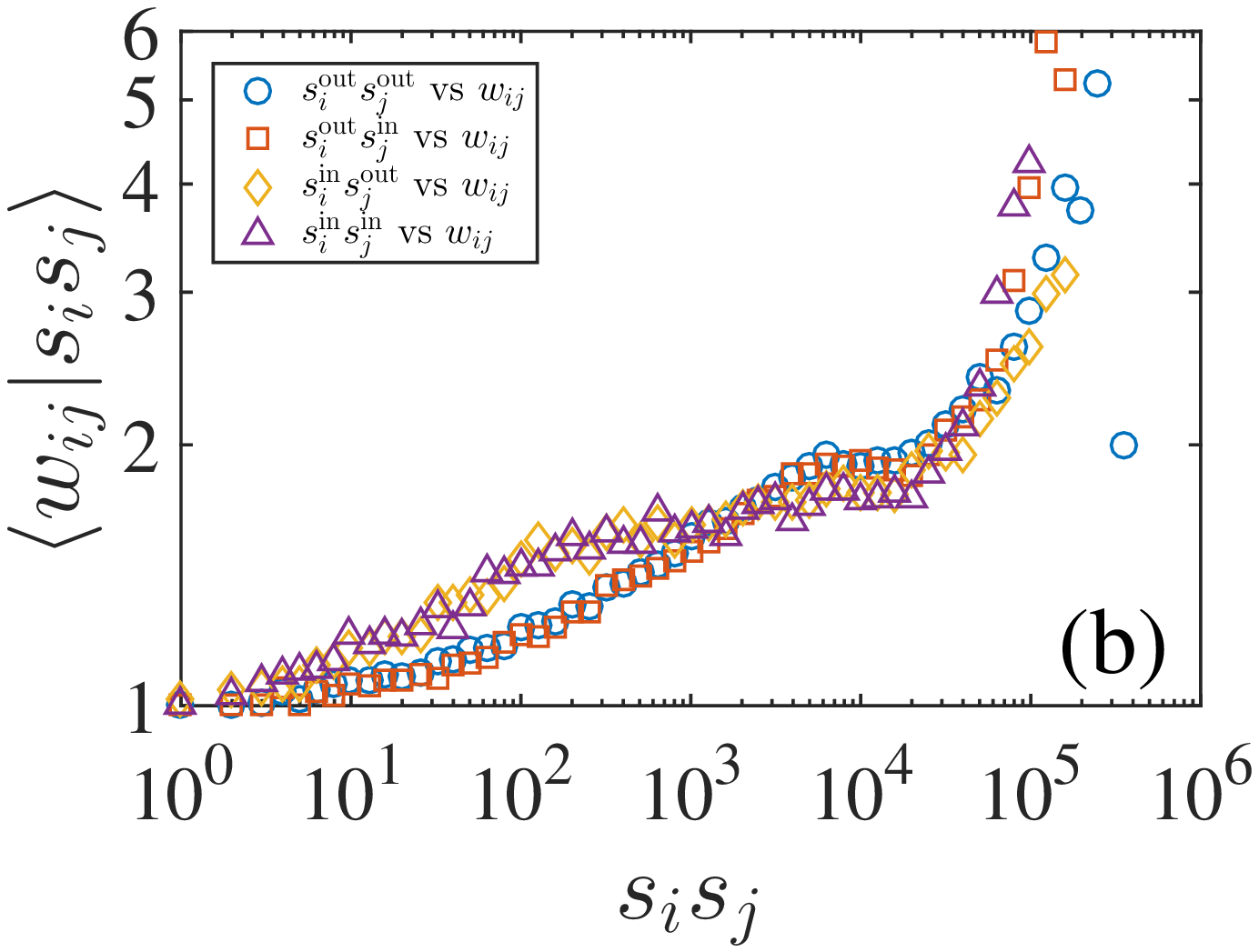}
  \caption{\label{fig:corr:3} Dependence of link weight $w$ with respect to node degree $k$ and node strength $s$. (a) The average link weight $\langle w_{ij} | k_ik_j\rangle$ of the directed links from $i$ to $j$ whose degrees are $k_i$ and $k_j$. (b) The average link weight $\langle w_{ij} | s_is_j\rangle$ of the directed links from node $i$ to node $j$  whose strengths are $s_i$ and $s_j$.}
\end{figure}

As the correlation exists between node degrees and link weights, we study the variation curve between link weights and the product of node degrees and the product of node strengths. The results are shown in fig.~\ref{fig:corr:3}. We observe a significant upward trend in both of the two plots. Furthermore, the upward trend can be simply separated into two parts. The separated points here ($k_ik_j^0 \approx 5,000$, $s_is_j^0 \approx 20,000$) are close to the product of the separated points of the bimodal distributions of node degrees and node strengths.

\subsection{Network Centrality}

Network centrality is a measure of how central a node locates in the network. Obviously, node degree is one of the centrality measures. Besides that, we also investigate other two commonly used centrality measures, betweenness centrality and closeness centrality. Betweenness centrality characterizes the importance of a node in network flow \cite{Freeman-1977-Sociometry,Brandes-2001-JMS}:
\begin{equation}
  b_j =\sum_{i\neq j \neq k} \frac{\sigma_{ik}(j)}{\sigma_{ik}},
\end{equation}
where $\sigma_{ik}$ is the number of shortest paths from node $i$ to node $k$ and $\sigma_{ik}(j)$ is the number of those paths that pass through node $j$. By dividing the number of pairs of nodes that do not contain node $j$, we can obtain the normalized betweenness centrality $b^n_j=b_j/[(N-1)(N-2)]$. If $b^n_j$ is close to 1, all paths in the network will pass through node $j$. In a connected network, a node's closeness is the average shortest path length between this node and other nodes \cite{Bavelas-1950-JAcoustSocA}. Since the FTN is not fully connected, we use the Harmonic centrality instead \cite{Marchiori-Latora-2000-PA}:
\begin{equation}
  H^\text{out}(i)=\sum_{i\neq j}\frac{1}{d(i,j)}, H^\text{in}(i)=\sum_{i\neq j}\frac{1}{d(j,i)},
\end{equation}
where $d(i,j)$ is the shortest path length from node $i$ to node $j$. By dividing $N-1$, we get the normalized Harmonic centrality $H^n = H/(N-1)$. The larger $H^n_i$ is, the closer node $i$ is to other nodes. If $H^n_i = 0$, node $i$ is disconnected with all of the other nodes, which means node $i$ is the isolated node. If $H^n_i = 1$, node $i$ is directly linked with all the other nodes. The graph is a star-like network and node $i$ is the center of the network.

\begin{figure}[htb]
  \centering
  \includegraphics[width=0.9\linewidth]{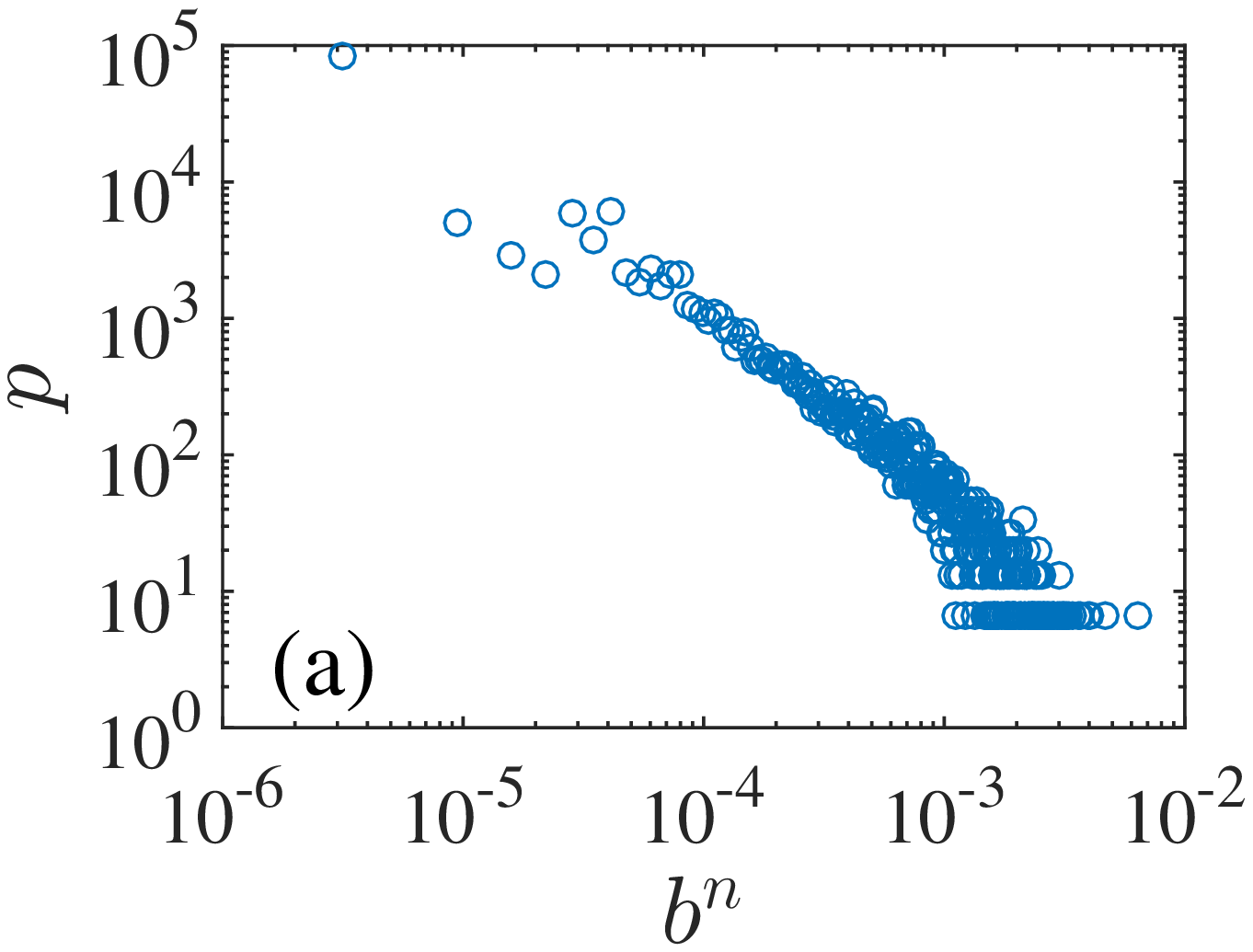}
  \includegraphics[width=0.9\linewidth]{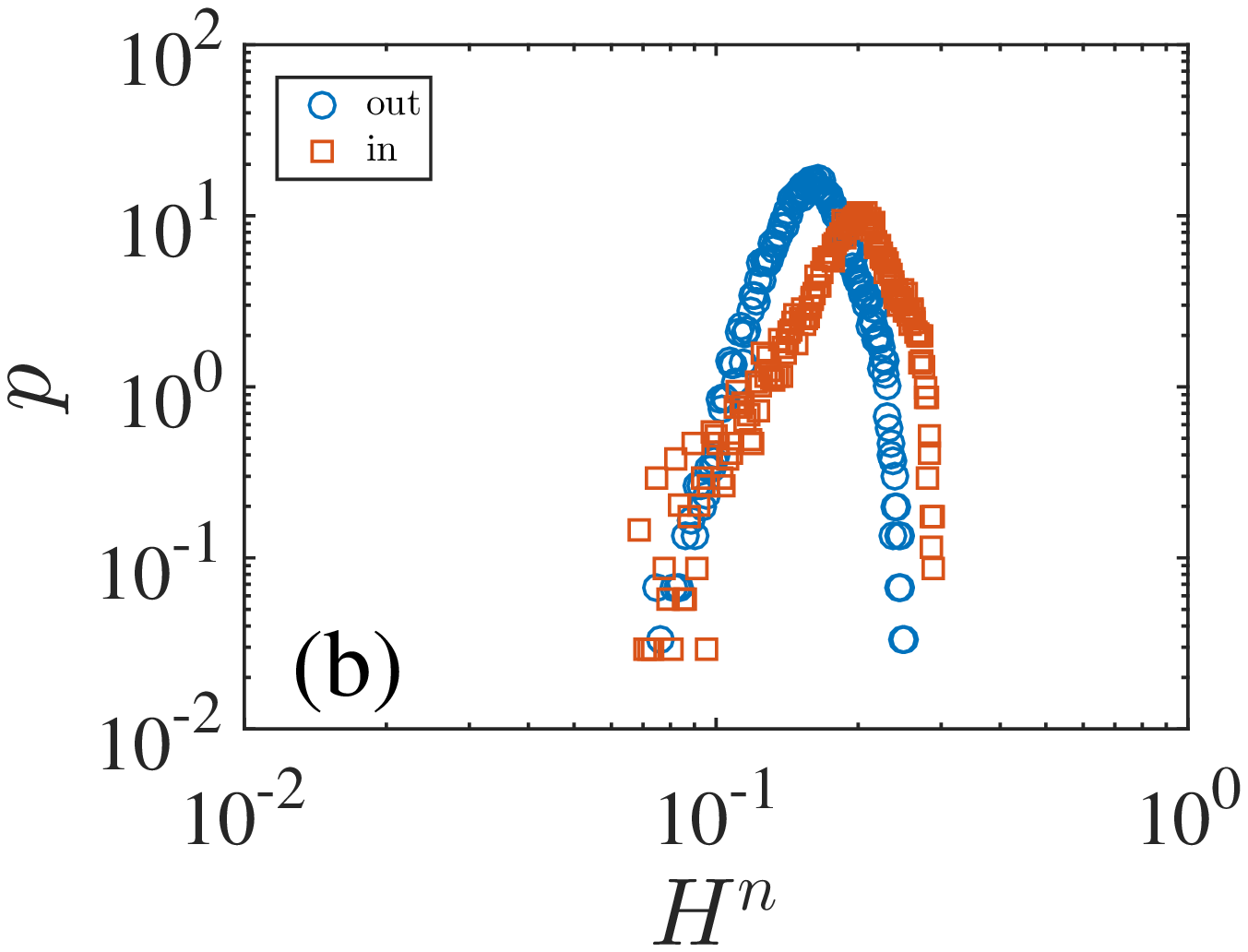}
  \caption{\label{fig:centrality:pdf} Empirical distributions of network centrality: (a) Betweenness centrality; (b) Closeness centrality. The nodes with $H^n < 0.001$ are not shown in the plot (2058 nodes for ``out'' and 6024 nodes for ``in''). Those nodes may be in the small connected components or close to the leaf nodes of the largest connected component.}
\end{figure}

Fig.~\ref{fig:centrality:pdf} shows the distributions of betweenness centrality and closeness centrality of the nodes in the FTN. One can observe that with the increase of $b^n$, the probability $p$ becomes smaller. The average normalized betweenness centrality $\langle b^n \rangle$ is close to 0.00011. It implies that most football clubs do not play an important role in the football players transfer market. The football club with the largest betweenness centrality ($b^n_{\max}= 0.0063$, about 600 times the average) is ``Parma (Italy)'', which is not surprising as ``Parma (Italy)'' has the largest in-degree and out-degree.
The distribution shape of the closeness centrality in fig.~\ref{fig:centrality:pdf}(b) is quite different from the betweenness centrality. One can find a peak presents in both curves. The peak locates at $H^n \approx 0.16$ for the closeness centrality from outgoing links and $H^n \approx 0.20$ for the closeness centrality from incoming links. And the distribution of $H^{n,\text{out}}$ is on the left side of the distribution of $H^{n,\text{in}}$. Considering the nodes with large closeness, we find that the closeness from outgoing links is smaller than the closeness from incoming links on average. Moreover, the number of nodes with small closeness $H^n < 0.001$ from incoming links is larger than that from outgoing links. It indicates that many leaf nodes only have outgoing links, corresponding to football training clubs.

In social network analysis, the centralization of a network is a measure of how equal the centralities of the nodes are in the network \cite{Freeman-1979-SN}, which is calculated as the differences between the centrality of the most central node and all other nodes. For a directed network, one considers centralization as the difference between the out-directional centralization and the in-directional centralization \cite{Li-Jiang-Xie-Xiong-Zhang-Zhou-2015-PA,Adamic-Brunetti-Harris-Kirilenko-2017-EmJ}, which is defined as follows \cite{Freeman-1979-SN},
\begin{equation}
\label{eq:centralization}
  C_{x}^{\rm{y}} = \frac{\sum\limits_{i=1}^N (x_{\max}^{\rm{y}} - x_i^{\rm{y}})}{\max \sum\limits_{i=1}^N (x_{\max}^{\rm{y}} - x_i^{\rm{y}})},
\end{equation}
where $x$ is one of the node centrality measures, $N$ is the number of nodes, $\rm{y}$ stands for superscript ``$\text{in}$'' or ``$\text{out}$'', and $\max \sum_{i=1}^N (x_{\max}^{\rm{y}} - x_i^{\rm{y}})$ is the maximal difference. Hence, $C_{x}^{\rm{y}}$ ranges from 0 to 1 and the centralization of a directed network $C_x = C^\text{in}_x - C^\text{out}_x$ ranges from $-1$ to $1$.

When considering an unweighted network, we can use node degree as the centrality measure, $C_k$, where $\max \sum_{i=1}^N (k_{\max}^{\rm{y}} - k_i^{\rm{y}}) = (N-1)^2$. If $C_k$ is close to $-1$, all the links in the network point from one node to the other nodes. If $C_k$ is close to $1$, all of the links point from other nodes to one node. When considering a weighted network, we calculate the strength centralization $C_s = C_{s}^{\rm{in}} - C_{s}^{\rm{out}}$, where $\max \sum_{i=1}^N (s_{\max}^{\rm{y}} - s_i^{\rm{y}}) = (N-1)\sum_{i=1}^N s^\text{y}_i$. The weighted network centralization  $C_s$ has the same physical meaning as the unweighted network centralization $C_k$. In the FTN, the degree centralizations are $C^\text{in}_k  = 0.0070$, $C^\text{out}_k  = 0.0095$, $C_k  = -0.0024$, and the strength centralizations are $C^\text{in}_s  = 0.0009$, $C^\text{out}_s  = 0.0017$, $C_s  = -0.0008$. All the centralizations are close to 0. It implies that no extreme dominating club exists in the global football player transfer market.

If we use $b^n$ for $x$ in Eq.~(\ref{eq:centralization}), we obtain the betweenness centralization $C_b$, where $\max \sum_{i=1}^N (b^n_{\max} - b^n_i) = N-1$. When $C_b$ is close to 1, some football clubs in the network act as the transfer center. Otherwise, there is no transfer center in the network. We obtain that $C_b = 0.0062$ for the FTN. If we replace $x$ in Eq.~(\ref{eq:centralization}) with $H^n$, we obtain the closeness centralization $C_H$, where $\max \sum_{i=1}^N (H^{n,\text{Y}}_{\max} - H^{n,\text{Y}}_i) = N-1$. For the FTN, we have $C^{\text{out}}_H = 0.1027$, $C^{\text{in}}_H= 0.1412$ and $C_H = 0.0385$. We find the closeness centralization from outgoing links is smaller than that from incoming links, which is different from the website network.

\section{Summary and discussions}

In this letter, we constructed a directed football player transfer network by using more than 470,000 transfer records of football players around the world from 1990 to 2016. We investigate the topological characteristics of the network.

We first investigated the distributions of in-degrees and out-degrees of node, link weights, and in-strengths and out-strengths of nodes. We found that the distributions of node degrees and node strengths can be fitted to bimodal distributions \cite{Wu-Zhou-Xiao-Kurths-Schellnhuber-2010-PNAS}, which might result from the different transfer paths of football players from different football clubs. We also found that the link weight distribution has a power-law tail.
We further inspected the correlations among node degrees, node strengths and link weights. It is found that the neighbor node degree (strength) increases with the increase of node degree (strength), indicating the FTN is assortative mixing. We also found that link weights correlate with node degrees, which is similar to the results in mobile phone communication network \cite{Li-Jiang-Xie-Micciche-Tumminello-Zhou-Mantegna-2014-SR}.
These properties have been studied for the mutual footballer transfer network \cite{Li-Xiao-Wang-Zhou-2018-IJMPB}. The corresponding properties for the FTN and MFTN are qualitatively similar, but they exhibit quantitative differences.

By analyzing various network centrality measures, we unveiled that most nodes have a small centrality, suggesting that no club acted as a transfer center. Investigation of the difference between the centralities of the most central node and all other nodes shows that all the football clubs occupy similar positions in the transfer network, and the closeness centralization from outgoing links is smaller than that from incoming links. It suggests that further studies on the FTN at meso-scales or micro-scales are required to uncover the different roles the clubs may play in the golobal transfer market.

\acknowledgments

This work was supported by the National Natural Science Foundation of China (11605062), the National Social Science Foundation of China (17AZD042) and the Fundamental Research Funds for the Central Universities (222201818006, 222201822009, 222201825010).


\end{document}